\documentclass{aa}
\usepackage{graphicx}
\usepackage{natbib}
\bibpunct[;]{(}{)}{;}{a}{}{;}
\begin{document}
\title{Death rate of massive stars at redshift $\sim 0.3$
\thanks{Based on observations collected at the European
Southern Observatory, Chile (ESO Programmes 62.H-0833,
63.H-0322, 64.H-0390, 67.D-0422, 68.D-0273, 69.D-0453).}
}
\author{E.~Cappellaro\inst{1}, M.~Riello\inst{2,3,4}, 
G.~Altavilla\inst{5}, M.~T.~Botticella\inst{1,6}\\
S.~Benetti\inst{3}, A.~Clocchiatti\inst{7}, 
J.~I.~Danziger\inst{8}, P.~Mazzali\inst{8}, 
A.~Pastorello\inst{9}, F.~Patat\inst{2},
M.~Salvo\inst{10} M.~Turatto\inst{3},
S.~Valenti\inst{1,11}}
\offprints{E. Cappellaro, \email{cappellaro@na.astro.it}}
\institute{
  INAF - Osservatorio Astronomico di Capodimonte,
  Salita Moiariello, 16, I-80131, Napoli, Italy\\
  \email{cappellaro@na.astro.it}
  \and
  European Southern Observatory, K. Schwarzschild Str. 2,
  85748 Garching, Germany
  \and
  INAF - Osservatorio Astronomico di Padova, Vicolo
  dell'Osservatorio, 5, I-35122, Padova, Italy
  \and
  Dipartimento di Astronomia - Universit\'a di Padova, Vicolo
  dell'Osservatorio, 2, I-35122, Padova, Italy
  \and
Departament d'Astronomia i Meteorologia, Universitat de Barcelona, Mart\'i i
Franqu\'es 1,  08028 Barcelona, Spain", 
\and
Osservatorio Astronomico di Collurania, via M. Maggini, I-64100 Teramo, Italy
Dipartimento di Scienze della Comunicazione, Universit\'a di Teramo, 
  viale Crucioli 122, I-64100 Teramo, Italy
\and
  Departamento de Astronom\'{\i}a y Astrof\'{\i}sica, Pontificia Universidad
Cat\'olica, Chile
  \and
  INAF - Osservatorio Astronomico di Trieste,
  via Tiepolo, 11, I-34131, Trieste, Italy
   \and
  INAF - Osservatorio Astronomico di Arcetri, Largo E. Fermi 5,I-50125
  Firenze, Italy
  \and
  Australian National University, Mount Stromlo Observatory,
  Cotter Road, Weston ACT 2611, Australia
  \and
  Dipartimento di Fisica - Universit\'a di Ferrara, via del Paradiso 12,
 I-44100 Ferrara, Italy
 }
\date{Received .../ Accepted ...}
\abstract{We report the first result of a supernova search program
designed to measure the evolution of the supernova rate with
redshift. To make the comparison with local rates more significant we
copied, as much as possible, the same computation recepies as
for the measurements of local rates. Moreover, we exploited the
multicolor images and the photometric redshift technique to
characterize the galaxy sample and accurately estimate the detection
efficiency.

Combining our data with the recently published meaurements of the SN
Ia rate at different redshifts, we derived the first, direct
measurement of the core collapse supernova rate at $z = 0.26$ as
$r_{cc} = 1.45^{+0.55}_{-0.45} \, h^2$\,SNu [$h$=H$_0$/75]. This is a
factor three {\bf ($\pm$ 50\%)} larger than the local estimate. The increase for a look
back time of "only" 2.8 Gyr is more rapid than predicted by most of
the published models of the SN rate evolution.  Core-collapse SN rates
measure the death rate of massive star and, because of the short time
scale of evolution, can be translated in a measurement of the ongoing
SFR. Assuming a Salpeter IMF and the standard scenario for
core-collapse progenitors we derived to an estimate of the star
formation rate at redshift $3.1^{+1.1}_{-1.0} \times 10^{-2} ~
h^3\mbox{M}_\odot\mbox{yr}^{-1}\mbox{Mpc}^{-3 }$ which compare very
well with a recent estimate based on the measurement of the $H\alpha$
luminosity density at the same redshift.

 }
\titlerunning{Death rate of massive stars at $z\simeq0.3$}
\authorrunning{Cappellaro et al.}
\maketitle
\section{Introduction}

Supernova rates represent a link between the evolution of individual
stars and that of stellar systems. In particular, the rate of type
II$\,+\,$Ib/c SNe measures the death rate for core-collapse (CC) of
young, massive stars and, because of the short time-scale of the
progenitor evolution \citep{heger}, directly reflects the on-going
star formation rate (SFR) in a given environment. On the contrary, the
rate of type Ia SNe, which result from long-lived, low mass binary
systems \citep{branch} reflects the long-term star formation
history. Owing to the very high intrinsic luminosity, SNe can be seen
at very large distances and hence measurements of SN rate evolution
with redshift can be used to trace the history of the SFR with cosmic
age.

Studies of the evolution of SN rates with resdhift are now strongly
unbalanced towards theory.  In recent years, many authors have
published predictions of the SN rate as a function of redshift based
on the SN progenitor scenarios and the modeling of the cosmic star
formation history
\citep*{MAD98,sadat,dahlen,yungelson,kobayashi,sullivan,matteucci}. Despite
the strong interest, observational estimates of
the SN rate at high redshift are still very scanty. The few published
measurements are based on SN searches aimed at using type Ia SNe as
cosmological probes \citep{PAIN96,PAIN02,TONRY} and, as a consequence,
they are strongly biased towards type Ia. To date, there is no
direct measurement of the evolution of core-collapse SN rates.

With the goal to fill this gap, we initiated a long term project
aimed to measure the rate evolution with redshift for all SN
types. In this paper we report the first results of this effort,
namely an estimate of the core-collapse SN rate at redshift $z\sim
0.3$.

To reduce the systematics in the comparison with the local rate, our
strategy we followed as closely as possible the same approach used
in that context \citep*{stat97,stat99}.  The process consists of four
steps: $i)$ SN candidate detection and, when possible, spectroscopic
classification (Sec. \ref{search}); $ii)$ characterization of the
galaxy sample through galaxy photometric redshifts; estimate of distances
and calibration of the absolute luminosities for each galaxy
(Sec. \ref{galaxy}) ; $iii)$ evaluation of the SN detection efficiency
and thus of the effective surveillance time for each galaxy of the
sample (Sec. \ref{efficiency}), $iv)$ estimate of the SN rate per unit
luminosity. To round up the paper, we discuss the main sources
of uncertainty (Sec.~6) and discuss the implication of our results
(Sec.~7).

\section{The supernova search}\label{search}

The building block of a SN search is the detection of variable sources
through the comparison of images of selected sky fields obtained at
different epochs. In general, the temporal sampling of the
observations is tuned to the specific goal one wants to achieve. For
the use of type Ia as cosmological distance indicators it is crucial
to catch SNe as early as possible and hence, accounting for the
typical rise time, the observations have to be spaced by 2-3 weeks.
Instead, to maximize the event statistics, the time elapsed between
exposures should be equal (or longer) than the time a typical event
remains brighter than the search detection limit. The latter, of
course, depends on the target distance (or redshift) and the SN type.
{\bf Independently on the temporal sampling, to ensure that all SNe are detected,
the time elapsed between the first and the last observation of a given field have to be
longer that time for a significant luminosity evolution for all SN type. The latter can be as long as 3-4 months for SN IIP and even longer for type IIn.}

For our search we selected 21 fields, evenly distributed in right
ascension, which have been monitored for about 2 years with an average
sampling of one observation every three months. However, the results
reported in this paper are based on observations of 5 fields only,
i.e. those with the best temporal and filter coverage. The full sample
will be presented in a future paper (Riello et al. 2004, in 
preparation). {\bf The observing log is shown in Tab.~\ref{tab:obs}.
For each field we give the center coordinates and, for the epochs when observations are available, we list the seeing. }

\begin{table*}
\caption{Observing log of the SN search fields. For each field, when observations in a given band are available, we report the measure of the seeing in arcsec (FWHM on the stellar objects on the combined image) }\label{tab:obs}
\begin{tabular}{lccccccccccccccc}
\hline
\hline
field           & \multicolumn{3}{c}{J1888} &
                  \multicolumn{3}{c}{AXAF}  &
                  \multicolumn{3}{c}{10Z2}  & 
                  \multicolumn{3}{c}{13Z3}  & 
                  \multicolumn{3}{c}{Field2} \\
R.A. (2000.0)   & \multicolumn{3}{r}{00h 57m 35.4s}  &
                  \multicolumn{3}{r}{03h 32m 23.7s}  & 
                  \multicolumn{3}{r}{10h 46m 45.8s}  & 
                  \multicolumn{3}{r}{13h 44m 28.3s}  & 
                  \multicolumn{3}{r}{19h 12m 51.9s}  \\
Dec.            & \multicolumn{3}{r}{$-$27d 39m 16s} &
                  \multicolumn{3}{r}{$-$27d 55m 52s} & 
                  \multicolumn{3}{r}{$-$00d 10m 03s} & 
                  \multicolumn{3}{r}{$-$00d 07m 47s} & 
                  \multicolumn{3}{r}{$-$64d 16m 31s} \\ 
\hline
  run            & B  & V & R  & B  & V & R & B & V & R &  B & V & R & B  & V & R  \\
\hline
1999/02/23       &    &   &    &    &   &   &1.1&   &   & 1.1&   &   &    &   &    \\
1999/03/10       &    &   &    &    &   &   &1.1&1.3&   & 1.2&1.2&   &    &   &    \\
1999/03/19       &    &   &    &    &   &   &1.0&0.9&   & 1.0&0.8&   &    &   &    \\
1999/05/08       &    &   &    &    &   &   &1.3&1.4&   & 1.4&   &   & 1.9&1.6&1.0 \\  
1999/05/17       &    &   &    &    &   &   &0.8&0.9&1.0& 1.7&1.0&0.9& 0.7&0.8&    \\    
1999/08/03       & 1.0&1.4&1.3 &    &   &   &   &   &   & 1.4&   &   & 1.1&1.8&    \\  
1999/09/13       & 1.6&1.4&1.2 &    &1.2&   &   &   &   &    &   &   &    &1.7&    \\   
1999/11/09       & 1.0&0.8&    & 1.1&1.0&   &   &   &   &    &   &   &    &   &    \\
1999/12/02       &    &1.0&    & 1.0&1.1&   &   &   &   &    &   &   &    &   &    \\
1999/12/10       &    &1.7&    &    &   &   &   &1.3&   &    &   &   &    &   &    \\
1999/12/28       &    &1.9&    &    &1.2&   &   &   &   &    &   &   &    &   &    \\
2000/11/16       &    &1.0&    &    &0.9&   &   &   &   &    &   &   &    &   &    \\
2000/12/17       &    &1.0&    &    &0.9&   &   &   &   &    &   &   &    &   &    \\
2001/04/18       &    &   &    &    &   &   &   &1.0&   &    &0.9&   &    &0.8&    \\  
2001/11/11-12    &    &0.7&1.2 &    &1.0&0.9&   &   &   &    &   &   &    &   &    \\
2001/11/18       &    &1.0&    &    &0.8&   &   &   &   &    &   &   &    &   &    \\
2001/12/08-09    &    &0.9&0.8 &    &1.0&0.8&   &   &   &    &   &   &    &   &    \\
2002/04/07-08    &    &   &    &    &   &   &   &1.9&0.9&    &0.8&1.1&    &1.3&1.3  \\
\hline
\end{tabular}
\end{table*}

A typical observing run was splitted into two parts: the search and the
follow-up observation of candidates.

For the search, two consecutive nights were devoted at the ESO/MPI
2.2m telescope at ESO, La Silla (Chile). The telescope was equipped
with the Wide Field Imager (WFI) and  a mosaic of $2
\times 4$ CCD detectors of $2048 \times 4096$ pixels which image a
sky area of $\sim 0.25$ deg$^2$ with an excellent spatial resolution
of 0.238 arcsec/pix. 

{\bf When possible}, the first observing night was dedicated to obtain deep $V$ band exposures for candidate detection while in the second night the same fields were observed through a
different filter, $B$ or $R$, with the purpose of collecting color
information both for the candidates and the galaxies. 
{\bf Unfortunately, due to a number of technical, meteorological and scheduling 
constraints in many cases we could not maintain this observing strategy as can be seen from Tab.~\ref{tab:obs}. This implies that only in a few cases we could derive the candidate color.
For homogeneity, in the following statistical computation we considered 
only the candidates detected in the V band exposures.} 

In order to get rid of detector cosmetic defects, cosmic rays, satellite tracks and
fast moving objects, for each field we obtained three 900 s exposures
dithered by 5-10 arcsec.

Follow-up observations were scheduled about one week after the search
at the VLT+FORS1/2 at ESO Paranal for the
spectroscopic classification of some of the candidates. The VLT was
needed, as most of our SN candidates are in the magnitude range
$V\simeq22.5$--$23.5$ mag. For a proper subtraction of the night sky
emissions we selected grisms of moderate resolution, namely grism 300V
and/or 300I (resolution $\sim10$\AA\/ FWHM), which allowed us to cover a
quite wide wavelength range ($4000\div11000$\AA). Depending on the
candidate magnitude, exposure times ranged from 900s to 3 hours.
Details on the reduction of the spectroscopic observations and of the
spectral analysis will be given elsewhere. Here we make use only of
the spectral classification and redshift.

The analysis of the search images began with the removal of the
instrument signature and calibration for which we used
IRAF\footnote{IRAF is distributed by the National Optical Astronomy
Observatories, which are operated by the Association of Universities
for Research in Astronomy, Inc., under cooperative agreement with the
National Science Foundation.} and {\em MSCRED}, a package specifically
designed to handle mosaic images \citep{mscred}.  Indeed, after bias
subtraction and flat fielding, the individual dithered exposures were
astrometrically calibrated and stacked in a single image. For
photometric nights, observations of standard fields were used for
absolute calibration \citep{landolt}. Otherwise, the photometric zero
point was established by comparison with a calibrated image of the
same field.

For each field we computed the difference between the image to be
searched (target image) and a suitable archive frame (template
image). Indeed, after accurate astrometric and photometric
registrations, the most crucial step in this process is the matching
of the point spread function (PSF) of the two images. This was done
using the {\em ISIS2.1} package \citep{isis} that, from the comparison
of the same sources in the two images, computes a space-varying
convolution kernel to degrade the image with the best seeing to match
the other one. Taking into account that the best subtraction is
obtained when the two images have similar PSFs, and that we want to
preserve as much as possible the resolution of the target image, we
had to maintain an archive with template images with different
seeing. Populating the archive required a significant investment of
telescope time and this was the reason why the search became really
efficient only after some time from the first observations.

\begin{table*}
\centering
\caption{SN candidates of the fields of Tab.~1 }\label{candidate}
\begin{tabular}{lccccccrcl}
\hline
\hline
designation &  R.A.       &Dec.   & \multicolumn{2}{c}{V mag} & offset & z     &  n.  & first  & search\\
            &\multicolumn{2}{c}{2000.0}       & cand. &  host & arcsec & phot. & exp. & detect.& class. \\
   \hline
J1888-H  & 00h 56m 26.25s & $-27$d 43m 31.8s  & 22.9  &  20.0 & 0.4    & 0.19     &  3 & 00/11/16& SN	  \\           
J1888-F  & 00h 56m 26.89s & $-27$d 27m 56.5s  & 22.5  &  16.1 & 0.4    & 0.20     &  6 & 99/11/09& SNAGN      \\ 
J1888-D  & 00h 56m 31.54s & $-27$d 31m 17.2s  & 23.5  &  21.1 & 0.4    & 0.30     &  6 & 01/11/11& SN	   \\      
J1888-C  & 00h 56m 33.40s & $-27$d 52m 52.3s  & 21.9  &  21.5 & 0.2    & 0.42     &  9 & 99/12/10& SN         \\   
J1888-L  & 00h 56m 38.43s & $-27$d 45m 07.9s  & 23.4  &  20.1 & 0.1    & 0.32     &  3 & 99/11/09& SNAGN      \\ 
SN2001gh & 00h 57m 03.63s & $-27$d 42m 32.9s  & 20.8  &  22.2 & 0.4    & 0.27     & 15 & 01/12/08& SN spec   \\ 
J1888-M  & 00h 57m 05.34s & $-27$d 45m 57.7s  & 23.4  &  21.8 & 0.2    & 0.19     &  3 & 01/12/08& SNAGN     \\         	
J1888-G  & 00h 57m 29.24s & $-27$d 40m 56.3s  & 22.6  &  19.0 & 0.2    & 0.17     &  3 & 99/12/10& SNAGN      \\       	 
J1888-B  & 00h 57m 48.08s & $-27$d 54m 13.3s  & 21.0  &  21.8 & 0.1    & 0.56     & 15 & 99/08/03& SN         \\ 
J1888-Q  & 00h 57m 51.11s & $-27$d 51m 28.3s  & 23.8  &  22.0 & 0.5    & 0.18     &  3 & 99/11/09& SN         \\
J1888-N  & 00h 58m 01.76s & $-27$d 53m 15.5s  & 23.5  &  21.0 & $<$0.1 & 0.32     &  6 & 99/11/09& SNAGN      \\ 
SN1999ey & 00h 58m 03.42s & $-27$d 40m 31.2s  & 20.2  &  21.3 & 1.1    & 0.01     & 15 & 99/11/09& SN spec   \\  
J1888-K  & 00h 58m 33.21s & $-27$d 27m 56.9s  & 23.3  &  20.5 & 0.2    & 0.31     &  3 & 99/12/02& SNAGN      \\       
SN2000fc & 00h 58m 33.55s & $-27$d 46m 40.1s  & 22.5  &  22.6 & 0.4    & 0.33     &  6 & 00/11/16& SN spec    \\  
J1888-J  & 00h 58m 41.36s & $-27$d 50m 38.1s  & 22.9  &  20.7 & 0.6    & 0.38     &  3 & 99/11/09& SN         \\           	
SN2001ip & 03h 31m 13.03s &  $-27$d 50m 55.5s & 23.5  & 21.8  &  0.1   & 0.39      & 6 & 01/12/08& SN spec    \\
AXAF-E   & 03h 31m 17.11s &  $-28$d 04m 47.9s & 23.3  & 21.8  &  0.1   & 0.24     & 3 &  00/12/17& SNAGN      \\
AXAF-A   & 03h 31m 28.46s & $-28$d 07m 57.4s  & 23.0  &  22.6 & 0.3    & --      &  3 &  99/12/28& SN         \\
AXAF-D   & 03h 31m 28.98s &  $-28$d 10m 26.1s & 23.4  & 22.8  &  0.2   & 0.52      & 6 & 99/12/28& SN         \\
AXAF-J   & 03h 31m 39.12s &  $-27$d 53m 00.5s & 23.7  & 18.9  &  0.2   & 0.14     & 3 &  00/12/17& SNAGN      \\                            
AXAF-H   & 03h 31m 49.98s & $-28$d 09m 41.6s  & 23.6  &  22.7 & 0.4    & --      &  3 &  91/12/08& SN         \\
AXAF-I   & 03h 32m 11.18s &  $-28$d 03m 49.6s & 23.7  & 20.3  &  0.2   & 0.15     & 3 &  01/12/08& SNAGN      \\                              
SN1999gt & 03h 32m 11.57s &  $-28$d 06m 16.2s & 22.0  & 20.8  &  2.2   & 0.17      & 3 & 99/12/28& SN spec    \\  
AXAF-B   & 03h 32m 31.16s &  $-28$d 04m 43.9s & 22.7  & 18.9  &  0.4   & 0.17      & 3 & 99/12/28& SN         \\
AXAF-C   & 03h 32m 45.62s &  $-28$d 08m 41.7s & 23.8  & 20.2  &  0.1   & 0.22     & 3 &  99/12/28& SNAGN      \\
SN1999gu & 03h 33m 00.22s &  $-27$d 51m 42.7s & 21.6  & 18.9  &  2.6   & 0.15      & 3 & 99/12/28& SN spec    \\  
AXAF-F   & 03h 33m 05.30s &  $-27$d 54m 09.2s & 23.2  & 21.8  &  0.2   & 0.26      & 3 & 99/12/02& SN         \\ 
SN2002cl & 13h 44m 09.94s &  $-00$d 12m 57.8s & 21.6  & 16.5  & 3.0    & 0.07     & 6 &  02/04/07& SN spec    \\
10Z2-B   & 10h 45m 42.76s &  $+00$d 00m 28.0s & 22.8  & 21.7  & 0.5    & 0.10     & 6 &  02/04/07& SN         \\       
10Z2-A   & 10h 47m 06.52s &  $+00$d 00m 39.7s & 21.9  & 21.1  & 0.3    & 0.69     & 6 &  02/04/08& SN         \\
10Z2-D   & 10h 47m 40.94s &  $-00$d 13m 52.7  & 23.4  & 21.5  & 0.6    & 0.76     & 6 &  99/03/19& SN  	   \\   
13Z3-A   & 13h 43m 28.86s &  $-00$d 14m 22.1s & 23.8  & 21.9  & 0.5    & 0.39     & 6 &  02/04/06& SN	   \\ 
13Z3-D   & 13h 43m 42.41s &  $+00$d 06m 44.9s & 23.2  & 20.9  & 0.2    & 0.46     & 6 &  99/03/10& SNAGN      \\      
13Z3-I   & 13h 44m 18.49s &  $-00$d 20m 55.5s & 23.7  & 21.8  & $<0.1$ & 0.13     & 3 &  99/03/19& SNAGN      \\      
13Z3-K   & 13h 45m 07.50s &  $+00$d 04m 06.9s & 23.4  & 20.1  & 0.2    & --       & 9 &  99/03/10& AGN spec   \\
13Z3-C   & 13h 45m 26.39s &  $-00$d 18m 10.6s & 23.1  & 19.8  & 0.7    & 0.15     & 3 &  01/04/18& SN	   \\        
13Z3-H   & 13h 45m 26.95s &  $+00$d 08m 04.0s & 23.5  & 20.7  & 0.3    & 0.28     & 3 &  99/05/17& SN         \\       
SN2001bd & 19h 13m 10.94s &  $-64$d 17m 07.8s & 21.7  & 16.8  &  3.5   & 0.15     & 3 &  01/04/18& SN spec    \\
Field2-F & 19h 14m 01.64s &  $-64$d 22m 38.6s & 22.7  & 22.0  & 0.8    & 0.33     & 3 &  01/04/18& SN	   \\  
Field2-G & 19h 14m 51.24s &  $-64$d 19m 34.9s & 22.9  & 20.9  & $<$0.1 & 0.07     & 3 &  01/04/18&  SNAGN     \\    
\hline
\end{tabular}
\end{table*}

Variable sources leave residuals in the difference image which have
been detected and logged into a catalogue using the {\em SExtractor}
program \citep{sex} which has also the capability of separating stars
from galaxies.  Due to residuals of poorly removed bright stars or
cosmic rays, the variable source catalogue contains many false
detections, most of which are quickly eliminated by means of a custom
made ranking program. This makes use of information from the
difference image as well as from the target and template images and it
has been tuned through extensive artificial stars experiments. The
surviving candidates, typically a few tens per field, are all checked
visually by a human expert.  Among these, a few are still obviuos
false detections which could not be properly flagged by our software
but are quickly eliminated by visual inspection.  Among these are
residuals of moving objects which are not completely masked by our
dithering strategy.  After that, we are left with true variable source
among which we remove variable stars, objects with stellar profile
present both in the target and template image, but with different
magnitudes. At the end of this process, one is left with the fiduciary
SN candidates, typically from none to a handful per field.

Ideally, one would need spectroscopic confirmation for all the
candidates. Unfortunately, because of the limited VLT time available,
we could obtain spectroscopic observations for $\sim20\%$ of the
detected SN candidates only. This is the main weakness of the work
presented here.  On the other hand, we could verify the reliability of
our SN candidate selection criteria: out of the 29 candidates for
which we have obtained VLT spectra during the entire search project,
22 turned out to be SNe (45\% type Ia and the other type II and Ib/c)
and 7 variable AGNs. We stress that our approach to candidate
selection was designed to avoid as much as possible any selection bias
and in particular, we do not exclude a priori nuclear
candidates. Given that, the intrusion of AGNs is unavoidable.

Even without spectroscopy, contamination by variable AGNs can be
reduced by looking at the long term variability history of the
candidates. With this aim we kept a database, powered by {\em MySQL}
\footnote{MySQL is an open source database without a license fee under
the GNU General Public License (GPL). See the project homepage for
further details. {\tt http://www.mysql.com/}}, which is used to search
for multiple detections of the same source \citep{searchdb}. If the
source shows long-term, erratic variability, it is excluded from the
list of SN candidates. For the candidates which passed this selection
and that were centered on the host galaxy nucleus ($\sim 30\%$) we
found that $40\%$ were actually SNe while the remaining ones were
still AGNs. To take this into account, in all the calculations nuclear
candidates were given a statistical weight of 0.4.  Note that with the
progressing of the monitoring, AGN contamination will continue to be
reduced.

In the 5 fields discussed in this paper, we have detected 40
candidates. {\bf These are listed in Tab.\ref{candidate} where we report the candidate designation (col. 1),
coordinates (cols. 2-3), and apparent V magnitude at discovery (col. 4), the apparent V magnitude of the host galaxy (col. 5), the offset (in arcsec) from the host nucleus (col. 6), the host photometric redshift (col. 7), the number of individual exposures in which the object has been seen (col. 8) and the epoch of first detection (col. 9). In the last column we report the classification of the candidate. Candidates located in the host galaxy nuclear regions are labelled SNAGN and are given a smaller weight as described above. }
 
For 9 of the candidate we obtained spectroscopic observations
which were used to derive spectral classification and redshift
(Tab. \ref{sne}). Eight of these objects were confirmed as SNe,
classified as type Ia (3), type II (4) and type Ic (1). One object was
found to be a Seyfert 1 galaxy.

\begin{table}
\centering
\caption{SN candidates with spectroscopic
classification}\label{sne}
\begin{tabular}{ccllc}
\hline
\hline
Designation & Type & Redshift & Field & Reference\\
\hline
 SN1999ey & IIn & 0.094 & J1888  & IAUC 7310\\
 SN1999gt & Ia  & 0.275 & AXAF   & IAUC 7346\\
 SN1999gu & II  & 0.149 & AXAF   & IAUC 7346\\
 SN2000fc & Ia  & 0.430 & J1888  & IAUC 7537\\
 SN2001bd & II-L & 0.096 & Field2 & IAUC 7615\\
 SN2001gh & II-P & 0.159 & J1888  & IAUC 7762\\
 SN2001ip & Ia  & 0.536 & AXAF   & IAUC 7780\\
 SN2002cl & Ic  & 0.072 & 13Z3   & IAUC 7785\\
\\
 13Z3-K   & Sy1 & 0.362 &  13Z3  & -- \\
\hline
\end{tabular}
\end{table}

\section{The Galaxy Sample}\label{galaxy}

The estimate of SN rates in the local Universe relies on the
characterization of the galaxy sample which have been searched
\citep{stat97}. For each galaxy one needs to know the distance, which
enters the computation of the surveillance time, and the integrated
luminosity, which is used as a normalization factor. Indeed, it has
been demonstrated that the SN rate scales with the size of the parent
stellar population as measured from the integrated blue luminosity
\citep{stat93}. For nearby galaxies the relevant information can be
readily retrieved from published catalogues, while this is not the
case when one goes to larger distances.

In an attempt to follow the same approach for our intermediate
redshift SN project, we exploited the $B$, $V$, $R$ images obtained
during the search to measure magnitudes and colors of the galaxies
detected in our fields and we used them to derive distances and
absolute luminosities through the SED fitting photometric redshift
technique \citep{hyperz}.

With this aim we selected from the image archive, for each field and band,
the exposures obtained under the best conditions, in particular those
with seeing $< 1\arcsec$, which were stacked together using the {\em
SWARP} package by E. Bertin (cf. http://terapix.iap.fr/). This
produces a sensible gain in the $S/N$ ratio: while the limiting
magnitude of a typical search image is $V\sim24.5$, the same for the
stacked image is $V\sim26$ ($3-\sigma$ point source).

From the $V$ images, which compared to the other bands benefit from
longer cumulative exposure times, we built the galaxy catalogue
including all sources with {\em SExtractor} stellarity index $\le 0.9$
\citep{arnouts}. A further selection is performed excluding galaxies
fainter than $R=21.8$. While this limit was originally chosen to
improve our confidence in the photometric redshifts (see next), it has
also the advantage that up to this magnitude the S/N is good enough to
guarantee a clean star-galaxy separation and a good photometric
accuracy. For these selected galaxies, $B$ and $R$ magnitudes were
eventually measured adopting the same aperture as defined in the $V$
image.

Photometric redshifts were estimated using the {\em hyper-$z$} code
\citep{hyperz}. This program searches for the best match between the
measured colors of galaxies and the values in a grid created from a
library of spectral energy distribution (SED) templates for different
redshifts. It has been shown that with a proper sampling of the SED,
even with broad band filters, galaxy redshifts can be measured with an
RMS error as small as $\sigma \sim 0.05$ \citep{hyperz}. In our case,
however, with observations in only three bands we cannot match this
level of accuracy. In particular, we note that one of the most
significant features in the galaxy SED, the 4000\AA\/ break, falls
redward of the $R$ band for redshifts larger than $z>0.8$. This means
that the redshifts derived for more distant galaxies are very
uncertain.

On the other side we notice that, because of the observing strategy
and limiting magnitude, the SNe discovered in our search are all at
$z<0.8$ with a peak of the distribution at $z\sim0.3$. To remove as
much as possible the contamination of distant galaxies erroneously
estimated at low redshift, we removed from the sample galaxies fainter
than $R=21.8 $. This roughly corresponds to the magnitude $M_*$ at a
redshift $z=0.8$ \citep{combo17}, where $M_*$ is a parameter of the
Schechter function \citep{schechter} which is used to fit the galaxy
luminosity function. With this choice, galaxies with redshift $\ge
0.8$ contribute about $20\%$ of the total sample luminosity.

At the same time, low luminosity galaxies at low redshift are removed
from the sample. However at $z\sim0.3$, the average redshift of our
search, they contribute to only 1/4 of the total luminosity and they
are expected to give a small contribution to the SN
productivity. Indeed we found that out of the 40 SN candidates, only 2
({\bf AXAF-H and AFAX-A}, which were not included in the computations) were discovered in
galaxies fainter than $R=21.8$.

\begin{figure}
\resizebox{\hsize}{!}{\includegraphics{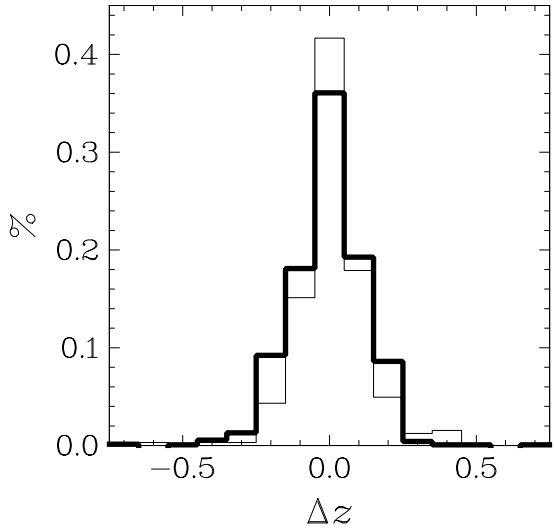}}
\caption{Distribution of $\Delta z = (z_{ph} - z_{sp})/(1+z_{sp})$,
the difference between our estimate of the photometric redshift and $a)$
spectroscopic redshifts, for the 324 galaxies of our sample with known
spectroscopic redshift (thin line) and $b)$ COMBO17 photometric
redshifts \cite{combo17} for the 1375 galaxies which are in common
with our sample (thick line)}
\label{ztest}
\end{figure}
 
The final galaxy catalog, which counts about 11300 galaxies,
was cross-checked with NED in order to assess the accuracy of the
photometric redshifts. We found that spectroscopic redshifts were
available for 324 galaxies (including 118 galaxies of the field J1888
for which spectroscopic redshift were kindly provided by P.-A. Duc, in
advance of publication).  With these data we built the histogram of
$\Delta z=(z_{\rm ph}-z_{\rm sp})/(1+z_{\rm sp})$, the differences between
spectroscopic and photometric redshifts, which is shown in Fig.
\ref{ztest} (dashed line). It turns out that the average difference is
$\langle\Delta z\rangle=0.01$ with a RMS error $\sigma=0.10$, which is
both consistent with our limited SED sampling and sufficient for our
statistical analysis.  An independent check on the accuracy of our
photometric redshift was made possible by the fact that one of our
fields (AXAF) partially overlaps with the Chandra Deep Field South
covered by the COMBO-17 survey \citep{combo17}. The distribution of
$\Delta z$ for 1375 galaxies which we have in common is also shown in
Fig \ref{ztest} (thick line). The average difference $\langle\Delta
z\rangle=0.00$ with a RMS error $\sigma=0.11$ is very similar to
those measured in the comparison with spectroscopic redshifts.

\begin{figure}
\resizebox{\hsize}{!}{\rotatebox{-90}{\includegraphics{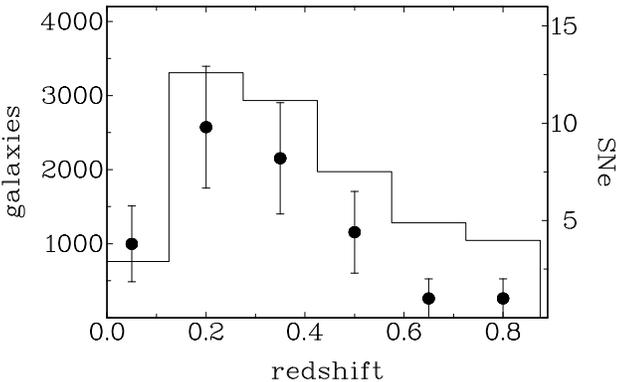}}}
\caption{Redshift distribution of our galaxy sample (line,
left-side scale) and SNe host galaxies (dots, right-side scale). For the
latter, candidates coincident with the host galaxy nucleus have been
counted with a 0.4 weight. }
\label{obs}
\end{figure}

The redshift distribution of our galaxy sample is shown in Fig.
\ref{obs}. As expected, given the adopted limiting magnitude, the
number counts peak at about $z\sim0.3$. For comparison, in
Fig. \ref{obs} we have also plotted the redshift distribution of the
SN candidate host galaxies (dots) with their statistical errors.

\section{Detection Efficiencies and Control Time}\label{efficiency}

The computation of rates requires the definition of the time interval
during which the events could be detected. We need to estimate the
time during which a SN hosted in a given galaxy remains brighter than the
search detection threshold which is usually referred to as {\em
control time}. This depends on $i)$ the search threshold and detection
efficiency; $ii)$ the SN absolute magnitude and luminosity evolution,
$iii)$ the distance.

There are several concurrent factors determining the detection
efficiency. For a given instrument and fixed exposure time, the key
factors are the observing conditions, in particular sky transparency
and seeing. There is also some dependence on the characteristics of
the particular sky field, due to the disturbing presence of bright
stars and nearby galaxies. The position of the candidate within the
parent galaxy plays also a role.

While the sky transparency of a specific observation is measured by
the photometric constants, the impact of other factors was evaluated
performing a number of artificial star experiments. In these
simulations we placed a number of synthetic stars with a given
magnitude $m$, and the PSF deduced from field stars, distributing them
in different galaxies of the field. The position of the artificial
stars within the host galaxy was chosen randomly, assuming a gaussian
distribution centered on the galaxy nucleus and a FWHM equal to that
of the host galaxy. Then, the synthetic frames were processed through
our search pipeline and the detection efficiency $\epsilon(m)$ was
computed as the ratio between recovered and injected stars.

\begin{figure}
\resizebox{\hsize}{!}{\includegraphics{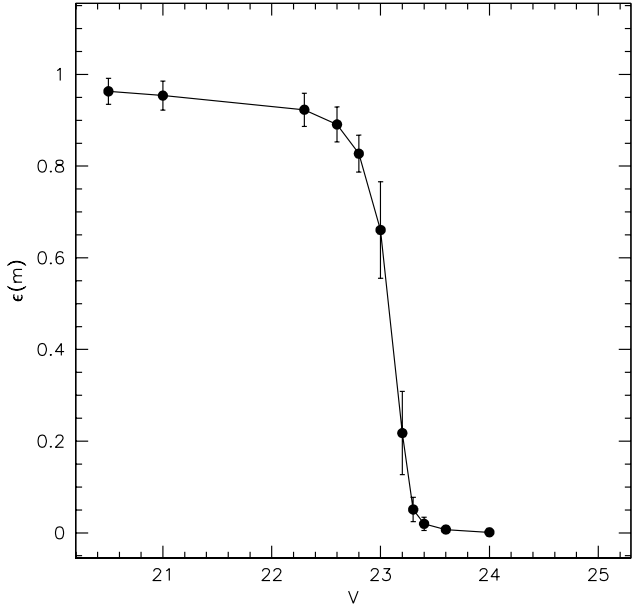}}
\caption{Example of SN detection efficiency curve, plotted as a function
  of apparent magnitude for one of the epochs of field AXAF (seeing 
  $0\farcs9$).
}\label{effs}
\end{figure}

In Fig. \ref{effs} we plot an example of detection efficiency function
for one of the observation of the field AXAF. As a result of these
numerical experiments, which will be described in detail in a
forthcoming paper (Riello et al. 2004), we found that the most
critical parameter is the seeing.  In particular we found that, taking
as reference the magnitude at which the detection efficiency is
$50\%$, this is $\sim1.2$ mag fainter when the seeing is 0\farcs65
compared with a seeing of 1\farcs3.

The other factor entering the computation of the control time for a
given epoch is the SN luminosity evolution and its absolute
luminosity.  The prescription we used in our calculations is described
in the following.  The apparent $V$ light curve for each SN was
computed taking into account the luminosity distance (estimated from
the photometric redshift of the host galaxy), the time dilation and
the $K$-correction according to the following relation:

\begin{equation}\label{lc}
  m^{\rm SN}_{i,V}(t) = M^{\rm SN}_B(t_0) +
  \mu(z_i) + K^{\rm SN}_{BV}(t_0) + A_V^G
\end{equation}

where, for the $i$-th galaxy at redshift $z_i$, $t_0 =t/(1+z_i)$ is
the galaxy rest frame time, $M^{\rm SN}_B(t_0)$ is the SN light curve
in $B$ absolute magnitude, $\mu(z_i)$ is the galaxy distance
modulus\footnote{Hereafter, we assume a standard flat cosmology
$\Omega_M=0.3$ and $\Omega_\Lambda=0.7$ and $h=H_0/75$ }, $K^{\rm
SN}_{BV} (t_0)$ is the $B$ to $V$ $K$--correction and $A^G_V$ is the
galactic extinction in $V$ (the upperscript {\it SN} indicates SN type
dependent quantities).

Following \citet{stat97}, we considered four basic SN types, namely
Ia, IIP, IIL and Ib/c. Guided by the general theoretical
interpretation, type IIP, IIL, and Ib/c will be collectively called
later {\em core collapse} (CC) SNe.  $B$ absolute magnitudes and light
curves were also taken as in \citet{stat97}. The $K^{\rm
SN}_{BV}$--correction for each SN type at different phases was
calculated using a sample of spectra with good $S/N$ ratio available
in the Asiago-Padova SN archive. Spectra were redshifted by the proper
amount and synthetic photometry was computed using the {\em synphot}
package in {\em IRAF}. The choice of the $K_{BV}$ correction is
motivated by the fact that, for the average redshift of our search,
$\langle z\rangle \sim0.3$, this minimizes both the uncertainties and
the time dependence of the $K$-correction. Note that for the same
reason we use as reference B light curves of local SNe.

We introduce here the effective control time defined, for each
observation of a given galaxy, as the control time weighted by the
galaxy luminosity in units of $10^{10}{\rm L}_{{\rm B}_{\sun}}$:

\begin{equation} 
CT^{\rm SN}_i = L_i 
\int{\tau_i^{\rm SN}(m)
\,\epsilon(m)\,{\rm d}m}
\end{equation}

where $\tau^{\rm SN}_i(m)$ is the time a SN in the $i$-galaxy
stays at a magnitude between $m$ and $m+{\rm d}m$, and can be computed
from relation (\ref{lc}), $\epsilon(m)$ is the detection efficiency at
the given magnitude and $L_i$ is the luminosity of the $i$-th galaxy
in units of $10^{10}$ blue--band solar luminosities.
 
Finally, for the given galaxy the total control time of the search
campaign, $\overline{CT}^{SN}_i$, is obtained by properly combining
the control time of individual observations \citep[cf.][]{stat99}.

\section{SN Rates}\label{rate}

For a given galaxy and SN type, the rates are derived by dividing the
number of observed events by the total control time:

\begin{equation} \label{eqrate}
r^{\rm SN}_i = (1+z_i) \frac{N^{\rm SN}}{\overline{CT}_i^{\rm SN}}
\end{equation}

where the factor $1+z$ corrects the rate to the rest frame.

The average redshift of the SN search $\langle z_{\rm SN}\rangle$ (i.e.
of the galaxy sample) is given by the average of the galaxy redshifts 
weighted by the effective control time:
\begin{equation} 
\langle{z}_{\rm SN}\rangle = \frac{\displaystyle\sum_{i=1}^N z_i \,
\overline{CT}_i^{\rm SN} }{\displaystyle\sum_{i=1}^N \overline{CT}_i^{\rm SN}}
\end{equation}
where N is the number of galaxies of the sample. It results that
$\langle{z}_{\rm Ia}\rangle=0.32$, $\langle{z}_{\rm cc}\rangle=0.26$ for Ia
and CC SNe respectively. The lower $\langle{z}\rangle$ for CC SNe is
obviously due to the fact that, on average, these are intrinsically
fainter than Ia.

Considering that we have a sample of galaxies with a wide spread in
redshift, to derive an estimate of the SN rates we have computed the
expected SN counts as a function of redshift, derived by summing the
contribution of individual SN types and using different assumptions
for the rate evolution with redshift. For the latter we assume a power
law dependence $r_{\rm SN}(z) = r_{\rm SN}^0 (1+z)^\alpha$, where
$r_{\rm SN}^0$ represent the local $(z=0)$ rate for a given SN type
and $\alpha$ is the evolution index ($\alpha=0$ indicates no
evolution). Using this model, the SN rate evolution parameters
can be derived from the best fit between the expected and observed
redshift distributions.

In practice, we derive the total expected SN detection $N^{\rm exp}(z)$
as the sum of
the expected number of Ia and CC events:
\begin{equation}
N^{\rm exp}(z) = N^{\rm exp}_{\rm Ia}(z) + N^{\rm exp}_{\rm cc}(z) 
\end{equation}
The expected redshift distributions of type Ia and CC SNe are
given by:
\begin{equation}
N^{\rm exp}_{Ia}(z) = \sum_{i=1}^n r^0_{\rm Ia} (1+z)^{\alpha_{\rm Ia}-1} 
\overline{CT}^{\rm Ia}_i(z) 
\end{equation}

\begin{equation}\label{eqcc}
N^{exp}_{\rm cc}(z) = \sum_{i=1}^n r^0_{\rm cc}  (1+z)^{\alpha_{\rm cc}-1} \overline{CT}^{\rm cc}_i(z)
\end{equation}
where the sums are extended over the $n$ galaxies in a 
given redshift bin $z$ and the effective control time for core 
collapse SNe is computed as follows:
\begin{equation}\label{eqctcc}
\overline{CT}^{\rm cc}_i(z) = f_{\rm Ib/c} \,\overline{CT}^{\rm Ib/c}_i(z) +  f_{\rm II}\, \overline{CT}^{\rm II}_i(z) 
\end{equation}
where we introduced the relative rates of type II, $f_{\rm II} = r^0_{\rm II}/
r^0_{\rm cc}$ and type Ib/c, $f_{\rm Ib/c}= 1-f_{\rm II}$. Here we make the further
assumption that the relative rates of different CC types do
not change with redshift and are equal to the local ones, i.e.
$f_{\rm II}=0.8$. \citep{stat99}.

As a first step, we have tested the null hypothesis that the SN rates
per unit blue luminosity do not change with redshift ($\alpha_{Ia}=
\alpha_{\rm cc} = 0)$ and they are equal to the local values, namely
$r^0_{\rm Ia} = 0.20\pm0.05$ [$h^2\,$SNu]\footnote{1 SNu = 1 supernova
century$^{-1}\,10^{-10}{\rm L}_{{\rm B}\sun}$.} and $r^0_{\rm cc}=0.47 \pm0.19$ [$h^2\,$SNu] \citep{stat99}.  Under this hypothesis, in our galaxy sample one would expect 13.2 events, which is significantly smaller than the 28.2
actually observed.  This already points out  that the global SN rate must
increase rapidly with redshift.

\begin{figure}
\resizebox{\hsize}{!}{\rotatebox{-90}{\includegraphics{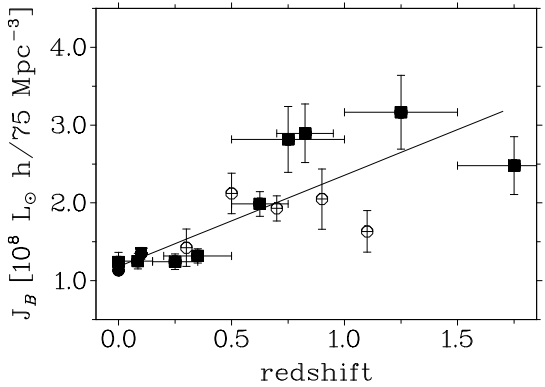}}}
\caption{Galaxy luminosity density ad different redshifts. Data are
  from \citep*{madaulb,combo17,blanton,norberg}. The line
  represents the best fit using a $(1+z)^{\alpha}$ power law (see the
  text for details).}
\label{lumden}
\end{figure}

In our case, since we do not have the spectroscopic classification for
all the SN candidates, we cannot directly use equation~(\ref{eqrate})
to derive individual SN rates.Unfortunately, with the limited
statistics available to date and the incomplete spectroscopic
classification, we cannot determine rates and evolutions of both SN Ia
and CC SNe from our data alone and we need to use other data sets. For
this purpose we have exploited the published measurements of SN Ia
rate at different redshifts \citep{blanc,sdss,hardin,PAIN02,TONRY} to
fix the evolution of SN Ia and to derive the CC rate.

One further complication is that some of the high redshift SN Ia
measurements were given only per unit volume
\citep{sdss,PAIN02,TONRY}.  To convert these numbers into rate per
unit luminosity (SNu scale), we need consistent estimates of the
luminosity density at different redshift. For this purpose we
collected from the literature recent measurements of the $B$
luminosity density for the redshift range relevant for our SN search
(Fig.  \ref{lumden}).  As can be seen, the measurements show a
relatively high dispersion, particularly large at high redshift.  The
best fit with a power law $\rho = \rho_0 (1+z)^{\alpha_\rho}$ gives
$\rho_0 = 1.18\times 10^8$
$[h\,\mbox{L}_{\mbox{B}\sun}\,\mbox{Mpc}^{-3}]$ and $\alpha_\rho =
1.0$. This has been used, in particular, to convert the rate per unit
volume of \citet{PAIN02} and \citet{TONRY}.

\begin{figure}
\resizebox{\hsize}{!}{\rotatebox{-90}{\includegraphics{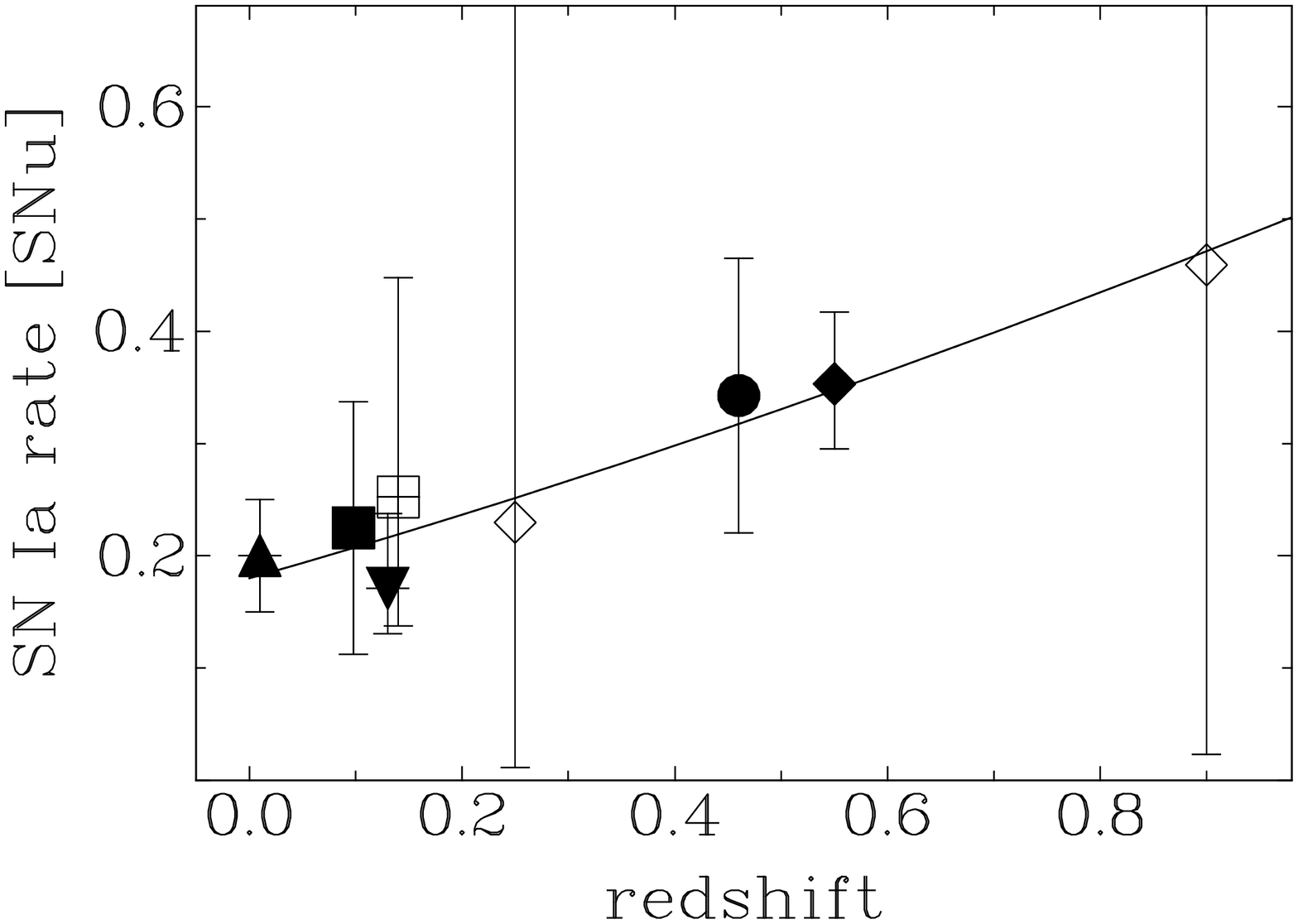}}}
\caption{Measurements of the SN Ia rate at different redshifts.
References are as follows: filled triangle - \citep{stat99}, filled
upside-down triangle - \citep{blanc}, filled square - \citep{sdss}, empty square -
\citep{hardin}, filled circle - \citep{TONRY}, filled diamond -
\citep{PAIN02}, empty diamond - \citep*{galyam}. The line is a fit with
the power law in $(1+z)$ (see the text for details). Because of the
large errorbars the Gal-Yam et al. estimates have not be used for the
fit.}\label{frateia}
\end{figure}

The published values of the SN Ia rate (in SNu) are plotted in
Fig. \ref{frateia}, which convincingly shows a redshift
evolution.  This can be fitted with a power law relation with
$r_{\rm Ia}^0=0.18\pm0.04$ $h^2\,$SNu and $\alpha_{\rm Ia} = 1.5\pm0.6$.

Assuming that the SN Ia rate evolution is known, we can use the
observed SN counts to constrain the evolution of core collapse rates.
To this end, it is convenient to re-write eq.~(\ref{eqcc}) as follows:

\begin{equation}
N^{\rm exp}_{\rm cc}(z) =\sum_{i=1}^n r^{\langle{z}_{\rm cc}\rangle}_{\rm cc}
\frac{(1+z)^{\alpha_{\rm cc}-1}}{(1+{\langle{z}_{\rm cc}\rangle})^{\alpha_{\rm cc}}}
\overline{CT}^{\rm cc}_i(z)
\end{equation}

which is referring the rate to the average redshift $\langle z_{cc}
\rangle=0.26$ in our case of the galaxy sample.

\begin{figure}
\resizebox{\hsize}{!}{\rotatebox{-90}{\includegraphics{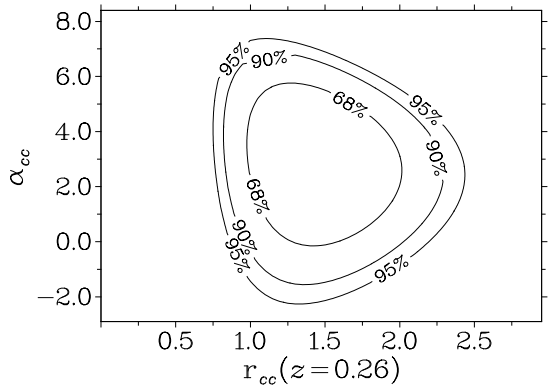}}}
\caption{Confidence level for the maximum-likelihood test.}
\label{fchisq}
\end{figure}

By means of a maximum-likelihood method we search for the values of
$r^{\langle{z}_{\rm cc}\rangle}_{\rm cc}$ and $\alpha_{\rm cc}$ which give the
best fit between observed and expected distributions as a function of
redshift. Resulting confidence levels are shown in Fig. \ref{fchisq}.
We found that $r_{\rm cc}(z=0.26)=1.45^{+0.55}_{-0.45}$ $h^2\,$SNu and
$\alpha_{\rm cc}=2.9^{+2.9}_{-2.9}$ where the quoted errors are the
1-$\sigma$ confidence level. Clearly the parameter $\alpha_{\rm cc}$
describing the CC-rate evolution is poorly confined by our data alone,
but the fair agreement with the measurement in the local Universe 
(cf. Fig. \ref{fratecc}) is conforting.

\begin{figure}
\resizebox{\hsize}{!}{\rotatebox{-90}{\includegraphics{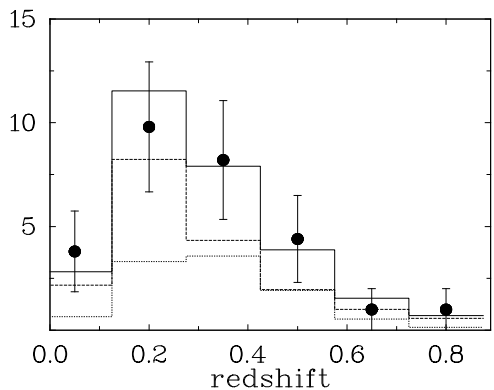}}}
\caption{Observed (points with statistical errorbars) and expected
(solid line) redshift distribution of SN counts. The short-dashed and
long-dashed lines are the expected type Ia and CC SN counts 
respectively.}
\label{fit}
\end{figure}

The comparison between the observed and expected distribution for the
best fit parameters is shown in Fig. \ref{fit}.  We notice that in our
search about 2/3 of the observed SNe are expected to be CC and only
1/3 SN Ia. This is consistent with what we have found when
spectroscopic confirmation was available (cf. Table~\ref{sne}).

The conclusion is that the CC SN rate at $\langle z\rangle\sim 0.3$ 
appears to be a factor 3 higher than in the local Universe.

\section{Uncertainties}

The errors quoted before for the rate and evolutionary index parameter
are purely statistical. Even though they are large, systematics errors
need also to be considered.  Indeed, although we made our best to
exploit the significant observational effort, this first estimate of
the CC rate at $z>0$ should be regarded as preliminary in many
respects.  In the following we highlight what we believe are the
most severe caveats. Note that we will not discuss here systematics
which originate from the uncertainties on the SN properties (absolute
magnitudes, light curves, intrinsic dispersion). For them, in fact, we
have made the same assumptions as for the computation of local rates
\citep{stat99} and therefore they are expected to cancel out in the
comparison. Also we do not address the uncertainties in the
cosmological model which was adopted and we rather focus on other aspects.

As we mentioned above, the most severe concern is the lack of
spectroscopic classification for all candidates. In particular, we had
to account in a statistical manner for the expected contamination by
AGNs, assigning a weight of 0.4 to the candidates coinciding with the
host galaxy nucleus . In order to evaluate the impact of this
assumption, we have computed the best fit for two extreme cases: {\em
a}) all nuclear candidates are AGNs or {\em b}) $80\%$ of the nuclear
candidates are indeed SNe. The two extreme cases encompass the ranges:
$r^{\langle{z}_{\rm cc}\rangle}_{\rm cc}=1.3\div1.9$ $h^2\,$SNu and
$\alpha_{\rm cc}=3.4\div2.3$ for the two fit parameters respectively.
Therefore although important, this uncertainty is not likely to affect
our conclusions.

Another concern is related to the limited accuracy in the
characterization of the galaxy sample through photometric redshifts,
due to the poor sampling of the galaxy SEDs. To check the possible
influence of this uncertainty, for the 1375 galaxies of the AXAF field
which we have in common with the COMBO17 surveys we computed the
expected number of events alternatively using our own estimates for
the galaxy redshifts and those reported in \cite{combo17}. We found
that to match the SN number count in the AXAF field, using as
reference the COMBO17 redshift estimates the rate $r^{\langle{z}_{\rm
cc}\rangle}_{\rm cc}$ has to be higher by $\sim 10\%$ compared with
our own redshift estimates.  This is a small difference which is
completely hidden by the statistical errors.

Of more concern is that in the current work we did not attempt to
correct for the bias due to extinction in the host galaxies. This
applies to all other estimates of SN rates at high redshift published
so far. It is well known that in local SN searches there is a severe
bias for SN detection in spiral galaxies which are not observed
face-on \citep{stat99}.  This is attributed to the concentration of
dust in the plane of the spiral galaxy disk which causes a higher
average extinction for the inclined ones. Because of the wavelength
dependence of extinction, it is expected that the blue photographic
surveys which were used to derive the estimates of the local rates are
heavily affected. For the same reason, it is sometimes claimed that
CCD searches in nearby galaxies do not need to be corrected. This
cannot be entirely true because the bias is seen even in the list of
events discovered in the last few years mainly by CCD SN searches
\citep{valencia}. But even if we accept this claim, we stress that if
high redshift galaxies have the same dust content as local ones then,
just because of redshift, host galaxy extinction should cause a similar
bias for blue band searches in nearby galaxies as for red
searches in high redshift galaxies. Unfortunately, current data on
high redshift SN searches do not allow to measure the size
of this bias.

At the moment, we can only add that any correction is likely to
increase, possibly even significantly, the CC estimate reported here.
In summary none of the reported uncertainties, even if important,
seems to undermine the main conclusion reached here, that a
significant evolution of the CC rate must be present even for this
small look back time.

\section{Discussion}

\begin{figure}
\resizebox{\hsize}{!}{\rotatebox{-90}{\includegraphics{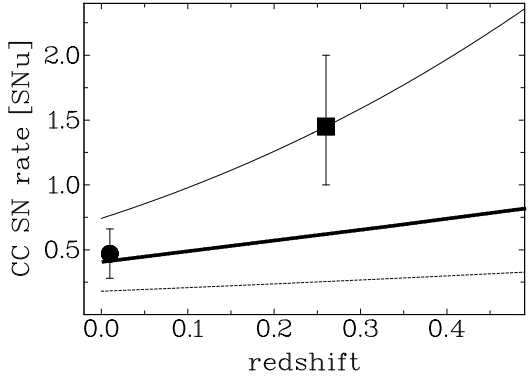}}}
\caption{CC SN rate with redshift. {\bf The dot is the estimate of the local CC SN rate from \cite{stat99} whereas the square is the new measurement derived in this paper.} The solid line shows the deduced
evolution with redshift. The dashed line shows the deduced type
Ia rate evolution (see the text for details). The thick solid line is
the CC SN rate evolution predicted by model M2 of \cite{sadat}. }
\label{fratecc}
\end{figure}

When we compare our result with the models we find that, in the
redshift range we are exploring, the current predictions indicate a
much shallower evolution of the CC rate
\citep{sadat,MAD98,matteucci,kobayashi} than actually observed..

As an example, in Fig.\ref{fratecc} we show the predictions 
 of \cite{sadat} in the most favorable case (model M2 which
corresponds to a higher SFR at high redshift). Although at a
2-$\sigma$ level the model is consistent with our measurement, we
remark that taken to face value the observed rate at z=0.26 is a factor
2 higher than the model.  This becomes even more significant if we
consider that, due to the lack of extinction correction, our estimate
is expected to be a lower limit to the actual CC rate.

On the other hand,  the CC rate evolution is directly
related to the adopted history of SFR for which there are many new
estimates. Indeed, considering the short evolutionary time scale of
the CC SNe progenitors ($< 5 \times 10^7$ yr) and under the assumption
that the initial mass function (IMF) and the mass range of the CC
progenitors do not change significantly in the redshift range of
interest, there is a simple direct relation between the SFR and the CC
rate, namely:

\begin{equation}
r_{\rm cc} = \psi \frac{\int_{M_l^{\rm cc}}^{M^{\rm cc}_u} \phi({\cal M}) d{\cal
M}}{\int_{M_L}^{M_U} {\cal M}\phi({\cal M}) d{\cal M}}$$
\end{equation}

where $\psi$ is the SFR at the given redshift, $\phi({\cal M})$ is the IMF,
$M_L-M_U$ is the mass range of the IMF and $M^{cc}_l-M^{cc}_u$ the
mass range for the CC SN progenitors.  In particular, adopting a
Salpeter IMF in the range $M_L=0.1$ to $M_U=125$ M$_\odot$ and
$M^{\rm cc}_l=8$, $M^{\rm cc}_u=50$ M$_\odot$ for the lower and upper limits
of the mass of CC SN progenitors we derive $r_{\rm cc} \simeq 0.007 \psi$.

Usually the cosmic SFR is reported per unit of comoving volume and
hence, for the comparison with SN rate evolution, it is convenient to
translate the latter to the same unit using the known evolution of the
luminosity density (Fig. \ref{lumden}). With this conversion, the CC
SN rate per unit volume results $r_{\rm cc}^V(z=0.26) = 2.2^{+0.8}_{-0.7}
\times 10^{-4} ~h^3 \mbox{yr}^{-1} \mbox{Mpc}^{-3}$ (whereas the local value
translates in $r^V_{\rm cc}(z=0) = 5.5 \pm 2.2 \times 10^{-5} ~h^3 \mbox{yr}^{-1}
\mbox{Mpc}^{-3}$).  Finally, using eq. 10, we converted the CC rate in SFR,
obtaining $\psi(z=0.26) = 3.1^{+1.1}_{-1.0} \times 10^{-2} ~ h^3
\mbox{M}_\odot \mbox{yr}^{-1} \mbox{Mpc}^{-3}$ (and $\psi(z=0) = 7.9\pm3.1 \times 10^{-3} ~
h^3 \mbox{M}_\odot \mbox{yr}^{-1} \mbox{Mpc}^{-3}$ for the local value).

\begin{figure}
\resizebox{\hsize}{!}{\includegraphics{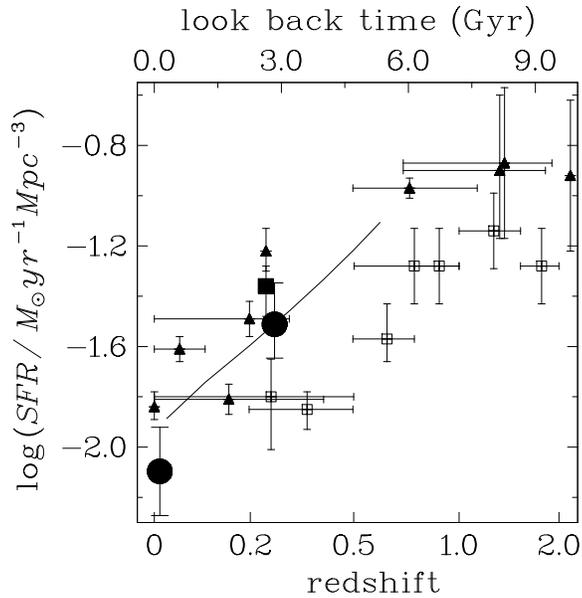}}
\caption{We compare our estimate of the SFR at redshift $z=0.26$ (big
filled dot) with the recent estimate of \cite{fujita} based on
the$H\alpha$ luminosity density at $z=0.24$ (filled square). Also
shown are estimates of the SFR at other redshifts based either on
measurements of the $H\alpha$ (filled symbols) or of the UV luminosity
density (adapted from \cite{fujita} and reference therein).  Also
plotted is the value derived from the local CC rate (also big filled
dot).}
\label{sfr}
\end{figure}

Recently, \cite{fujita} have published an estimate of the SFR based on
the $H\alpha$ luminosity density at $z\simeq0.24$.  This is shown in
Fig.\ref{sfr} where are also reported other estimates of the SFR at
different redshifts (adapted from \cite{fujita}, see references
therein).  As noted by \cite{fujita} there is a systematic difference
in the SFR deduced from the $H\alpha$ luminosity density (filled
symbols) compared with that derived from the UV luminosity density non
corrected for extinction (empty symbols), with the former being
significantly smaller at all redshift. However, it has been shown that
this disagreement can be removed with a proper extinction correction
\citep{terlevich,hippelein}.

As seen in Fig.~9, our measurement is in excellent agreement with the
value of \cite{fujita} and in general with the rapid SFR evolution
deduced from $H\alpha$ luminosity density.  Our conclusion, from the
evolution of the core collapse SN rate is that at at a redshift
$z=0.26$, that is at a look back time of 2.8 Gyr with the adopted
cosmology, the SFR per unit comoving volume was three time higher than
in local Universe. {\bf We stress again that, because our measurement of the CC SN rate 
is not corrected for extinction, this is likely to be a lower limit (cf Sect. 6).} \footnote{{\bf After submission of 
this article, two preprints have been posted reporting new estimates of the SN rate evolution up to redshift 1.6, as part of the  Great Observatories Origins Deep Survey
\citep{goods1,goods2}. The main caveat is that, like in our case, they could obtain spectroscopic classification only for a fraction of the candidates. In any case,
it turns out that their estimate of the CC SN rate at redshift z=0.3 is in remarkable agreement with the result reported here.}}

\begin{acknowledgements}
We like to thank Michele Massarotti for the advices on the photometric
redshift technique.  This research has made use of the NASA/IPAC
Extragalactic Database (NED) which is operated by the Jet Propulsion
Laboratory, California Institute of Technology, under contract with
the National Aeronautics and Space Administration.  AC acknowledges
support from CONICYT, Chile, through grant FONDECYT 1000524. SV
acknowledges support from the program ``Promozione della ricerca
scientifica Regione Campania'', Legge Regionale n. 5 del 28/03/2002.
\end{acknowledgements}

\end{document}